\documentstyle [12pt,aaspp4] {article}

 1

\def\ref{\parskip=0pt\par\noindent\hangindent\parindent
    \parskip =2ex plus .5ex minus .1ex}
\def\gs{\mathrel{\raise1.16pt\hbox{$>$}\kern-7.0pt
\lower3.06pt\hbox{{$\scriptstyle \sim$}}}}
\def\ls{\mathrel{\raise1.16pt\hbox{$<$}\kern-7.0pt
\lower3.06pt\hbox{{$\scriptstyle \sim$}}}}
\def\gtsima{$\; \buildrel > \over \sim \;$}
\def\ltsima{$\; \buildrel < \over \sim \;$}
\def\prosima{$\; \buildrel \propto \over \sim \;$}
\def\gsim{\lower.5ex\hbox{\gtsima}}
\def\lsim{\lower.5ex\hbox{\ltsima}}
\def\simgt{\lower.5ex\hbox{\gtsima}}
\def\simlt{\lower.5ex\hbox{\ltsima}}
\def\simpr{\lower.5ex\hbox{\prosima}}

\begin{document}
\title {Constraints on the Cosmic Structure Formation Models\\
from Early Formation of Giant Galaxies}
\author {H. J. Mo$ ^{1,2}$ and M. Fukugita$ ^{1,3}$}
\affil {$ ^1$Institute for Advanced Study, Princeton NJ 08540}
\affil {$ ^2$Max-Planck-Institut f\"ur Astrophysik, 
85748 Garching, Germany}
\affil {$ ^3$Yukawa Institute, Kyoto University, Kyoto, 606 Japan}
\authoraddr {Max-Planck-Institut f\"ur Astrophysik,
Karl-Schwarzschild-Strasse 1, 85748 Garching, Germany}
\slugcomment{submitted to ApJL}
\received{         }
\accepted{         }
\begin{abstract} 
A recent observation of Steidel et al. indicates that a substantial
fraction of giant galaxies were formed at an epoch as early as 
redshift $z>3-3.5$. We show that this early formation of giant galaxies
gives strong constraints on models of cosmic structure formation. 
Adopting the COBE normalization for the density perturbation spectrum,
we argue that the following models do not have large enough power on 
galactic scales to yield the observed abundance: 
(i) standard cold dark matter (CDM) models (where mass density 
$\Omega_0=1$ and power index $n=1$) with the Hubble constant 
$h\lsim 0.35$; (ii) tilted CDM models with $h=0.5$ and 
$n\lsim 0.75$; (iii) open CDM models with $h\lsim 0.8$ and
$\Omega_0 \lsim 0.3$, and (iv) mixed dark matter models with 
$h=0.5$ and $\Omega _\nu \gsim 0.2$. Flat CDM models with
a cosmological constant $\lambda_0 \sim 0.7$ are consistent
with the observation, provided that $h\gsim 0.6$. 
Combined with constraints from large-scale structure
formation, these results imply that the flat CDM model with a low
$\Omega_0$ is the only one that is fully consistent with observations.
We predict that these high-redshift galaxies are more 
strongly clustered than normal galaxies observed today.
\end{abstract}
\keywords {galaxies: formation-cosmology: theory-dark matter}

\section {INTRODUCTION}

 Recently Steidel and collaborators (Steidel \& Hamilton 1993; 
Steidel et al. 1996, hereafter S96)
have developed a novel technique
to detect high redshift galaxies using Lyman continuum break redshifted
to the wavelength range between $U$ and $G$ pass bands. They have
found a number of candidate galaxies having redshift $z=3-3.5$
in broad-band photometry, and their follow-up spectroscopy has 
confirmed that these galaxies indeed have such
redshifts or at least are consistent with having such redshifts. This
observation indicates that about 2\% of the galaxies in the magnitude 
range ${\cal R}_{AB}=23.5-25$ mag have redshift $z=3-3.5$; this means that
a substantial fraction ($\gsim 10-30$\%) of giant galaxies observed 
today have already been formed before this redshift. The observed 
spectrum and colors suggest that the formation epoch of these galaxies
could probably be earlier by $\Delta t\simeq$ 1Gyr.
S96 have also argued from the equivalent widths of saturated 
absorption lines that the velocity dispersion of these
galaxies is probably as high as 180-320 km s$^{-1}$, comparable to
that for $L>L^*$ elliptical galaxies observed today,
although the possibility that the dominant 
part of equivalent widths is caused by P Cygni profile of
gas outflows is not excluded.
While we have to await the confirmation from future
high resolution spectroscopy, it is very likely that they are
beginning to observe the early stage of spheroids of 
giant galaxies.

We note that this abundance information of high redshift galaxies 
gives a strong constraint on the model of cosmic structure
formation.  The current structure formation models are always 
tuned so that they yield successful predictions for the large-scale 
structure observed at present time. As a result it is difficult 
to discriminate them by using the information of large scale structure 
at low redshift alone.  On the other hand, the predictions for 
small-scale structure at an early epoch, such as high-redshift
galaxies, are quite different from 
model to model. We argue that even the present rather premature data 
on high redshift galaxies can discriminate models, 
provided that some extra information concerning the 
normalization of the primordial spectrum is given. 

We also show that the models that satisfy the test of the 
abundance of high-redshift galaxies predict a high
degree of bias of these galaxies relative to the mass 
compared to that of normal galaxies observed today. 

\section {MODELS}
 We take as a basis of our argument the Press-Schechter formalism
(Press \& Schechter 1974, hereafter PS), which allows an analytic 
treatment of the problem. The PS formalism has been tested 
extensively by $N$-body simulations for a variety
of hierarchical clustering processes in various cosmic 
structure formation models
(e.g. Bond et al. 1991; Bower 1991; Lacey \& Cole 1994; Mo \& White 1996;
Mo, Jing \& White 1996). In the PS formalism, the comoving number
density of dark halos in a unit interval of halo velocity 
dispersion $\sigma$ is given by
$$
{dN\over d\sigma}(\sigma,z)={-3\over (2\pi)^{3/2}}
{1\over r_0^3\sigma}{\delta_c(z)\over \Delta (r_0)}
{d\ln \Delta(r_0) \over d\ln r_0}
\left\lbrack {d\ln\sigma\over d\ln r_0}\right\rbrack^{-1}
\exp\left\lbrack -{\delta_c^2(z)\over 2\Delta^2(r_0)}\right\rbrack,
\eqno(1)
$$
where $r_0$ is the radius of a sphere that comprises a halo of mass $M$ 
for a homogeneous universe with mean mass density $\rho_0$, i.e.,
$M=4\pi\rho_0 r_0^3/3$;
$\Delta (r_0)$ is the rms of the linear mass density 
fluctuations in top-hat windows of radius $r_0$;
$\delta_c(z)$ is the critical overdensity for collapse
at redshift $z$.  The quantity $\Delta (r_0)$ is completely 
determined by the initial density spectrum
$P(k)$ (which is assumed to be Gaussian) and we normalize $P(k)$ 
by specifying $\sigma_8\equiv \Delta (8h^{-1}{\rm Mpc})$, where 
$h$ is the Hubble constant in units of 100 km s$^{-1}$Mpc$^{-1}$. 
For any cosmological model and power spectrum, one can 
calculate $dN/d\sigma$ given
the function $\delta_c(z)$ and the relationship between 
$\sigma$ and $r_0$. We take the result summarized by 
Kochanek (1995; see also Bartelmann et al. 1993) for these relations. 
As one can see from eq.(1), a change in $\delta_c(z)$ leads
to a proportional change in $\sigma_8$.   

  We consider five sets of cosmic structure formation models, each
containing one free parameter for which a constraint is to be derived. 
The first set is the standard CDM model 
(where the cosmic density of matter $\Omega_0=1$ and the power index 
of primordial density perturbation spectrum $n=1$) 
with $h$ a free parameter.  
The second set is the tilted CDM model with $h$ fixed
but with $n$ varying. The third one is a CDM model
in an open universe (open CDM), with $\Omega_0$ a free parameter.
In the forth set, we consider CDM model in a flat universe
(flat CDM: i.e. $\lambda_0+\Omega_0=1$ where
$\lambda_0\equiv\Lambda/3H_0^2$ is the cosmic density
parameter of cosmological constant), with $\Omega_0$ a free
parameter. Finally, in the fifth set, 
we discuss mixed dark matter (MDM) model (where $\Omega_0=1$)
with neutrino mass density $\Omega_\nu$ as a free parameter.
The model parameters are summarized in Table 1. 

For CDM models, the power spectra are calculated using the 
fitting formulae of Hu \& Sugiyama (1996). The contribution
of the gravitational wave to the normalization is ignored. 
We assume the baryon density parameter to be $\Omega_b=0.0125\, h^{-2}$ 
from primordial nucleosynthesis calculations (Walker et al. 1991). 
The amplitudes of the power spectra are estimated from the 4-year 
COBE data (Bennett et al. 1996) with the aid of the fitting 
formulae given by White \& Scott (1996). 
For the MDM model, we use the fitting formulae of Ma (1996) to 
estimate both power spectra and COBE normalizations. 

In the calculation of the comoving
number density of halos in eq.(1) we need to specify the velocity 
dispersion of galactic halos and the epoch when these halos are formed.
From the equivalent widths of heavily saturated lines, S96 have estimated 
the velocity dispersions of Lyman break galaxies to be $\sigma=$ 180-320
km s$^{-1}$. These values are compared to 200-230 km s$^{-1}$ 
for $L^*$ galaxies of E/S0 type.  Together with other indications, 
S96 concluded that they are indeed observing early phase of the spheroids
of $>L^*$ galaxies, although they did not exclude the
possibility that this high velocity dispersion is caused by outflow
rather than by gravitational motion.
Here we accept their interpretation that
this velocity dispersion is gravitational, and take the threshold
velocity dispersion to be $\sigma_{\rm min}=180$ km s$^{-1}$.
We note the possibility that the true halo velocity dispersion 
may be higher than that of stars (e.g., Gott 1977).
 
We evaluate the comoving number density of galaxies 
for two epochs: (i) 1 Gyr before the epoch
that corresponds to $z=3$, and (ii) at $z=3.5$.  Case (i) 
is probably a more realistic formation epoch for these galaxies,
as inferred from the ${\cal R}-G$ color assuming that some star
formation activity persists to $z=3$. Case (ii) is true only when
star burst is instantaneous: a strong rest-frame 
UV light implies that the burst epoch
is only 0.01 Gyr back from the observed epoch.
As noted by S96, this is an unlikely case, since it requires all 
observed galaxies to undergo completely coeval burst phase at the 
observed redshift. 
We take case (ii) as the most conservative estimate.

\section {RESULTS}

 The abundance estimate of Lyman break galaxies depends on assumed 
cosmology. S96 estimated that the comoving number density of 
these galaxies is $N_g\approx 2.9\times 10^{-3}h^{3}$ Mpc$^{-3}$ in the 
Einstein-de Sitter universe and $5.4\times 10^{-4}h^{3}$ Mpc$^{-3}$
in an open universe with $\Omega_0=0.1$. These abundances correspond
to 1/2 and 1/10 of the space density of present-day galaxies with 
$L>L^*$, respectively. Shimasaku \& Fukugita (1996) have 
argued, using their evolution model where all spheroidal components 
are assumed to have formed very early and passively evolved and disk 
components are added later with e-fold time of 5 Gyr,
that S96 are observing basically all E/S0 galaxies at this
redshift if the universe is open or $\Lambda$-dominated, and 
that the observed fraction is about 1/3 of what is expected 
if $\Omega_0=1$. For our argument here, it suffices to take the value of 
$N_g$ given by S96. For cosmologies other than those used in S96,
we estimate the density by modifying the comoving
volume in the redshift range $3.0\le z\le 3.5$ into 
the one for the relevant case. 

We should note that what is calculated with eq. (1) is the halo
abundance, and that some halos may not contain ``galaxies'' if
star formation is for some reasons inhibited in them.
The calculation of Shimasaku \& Fukugita indicates that this is unlikely
to happen at least for a low density or a $\Lambda$-dominated
universe. Taking into account the fact that the velocity dispersion of a
halo could be higher than that is observed for stars 
(see discussion in Section 2) and the fact
that some massive halos which do not contain Lyman break galaxies
may be missed, we take $N_g$ of S96 as a lower limit to the halo 
number density calculated with eq. (1).  

It is also possible that some massive halos contain 
more than one galaxies, and the number density of galaxies 
could be larger than that of the dark halos with velocity dispersion
bigger than $\sigma_{\rm min}$. This possibility is,
however, unlikely in the general field, as observed in S96
\footnote{We remark that our result would change little, even if
we allow for this possibility: at high 
redshift the total mass contained in ``massive halos'' is small.
We obtain basically the same result 
if we take the number density of galaxies to be the ratio between the 
total density of mass contained in dark halos 
with $\sigma>\sigma_{\rm min}$ and the typical mass implied by the 
observed velocity dispersion.}.

  Before considering individual models, let us examine the 
general sensitivity of our calculation to the parameters discussed
above. We plot in Figure 1 the abundances of `galaxies' predicted in 
the standard CDM model and in a flat CDM model with $\lambda_0=0.9$
as a function of the normalization $\sigma_8$. 
For each model, results are shown for four cases: 
two cases refer to two extreme values of $h$,
and one case where $\Omega_b$ is set to zero for the lower $h$ case. 
The other curve shows a somewaht different calculation, where
the number density of galaxies is  obtained by dividing the mean
density of mass contained in halos with $\sigma\ge\sigma_{\rm min}$
by a mass corresponding to $\sigma_{\rm min}$. The difference
among four curves are not large.
The change of $h$
has two compensating effects: an increase of $h$ 
pushes the redshift (for a given $\Delta t$) of halo formation
to a higher value and causes $N$ to decrease, but at the same time
enhances the power on small scales for
a given $\sigma_8$ [because $\Delta (r_0)$ becomes steeper] 
so as to increase the number density.
The net dependence of $N$ on $h$ is weak for the ranges
of $h$ relevant to our discussion. 
The inclusion of the baryonic component suppresses the power on 
small scales (see e.g. Hu \& Sugiyama 1996) and thereby reduces the 
number density of halos predicted for a given $\sigma_8$. 
The figure shows, however, that the change is small even for
low values of $h$ (and low $\Omega_0$), where the effect is 
expected to be stronger.
The calculation using halo mass also agree with other calculations
from equation (1), except for large 
$\sigma_8$ where a considerable amount of mass is already 
in large halos. 
For the range of $\sigma_8$ relevant to our discussion, however, 
the difference can be neglected.
Fig. 1 shows that, in any case, the values
of $\sigma_8$ required to give the observed abundance
do not depend sensitively on the detail of the calculation,
since the predicted curves crosses the observed $N_g$
(shown as the horizontal lines)  
where $N$ increases sharply with $\sigma_8$.   

  Our main results are summarized in Figure 2, 
which shows the values of $\sigma_8$ required to
give the observed comoving number density of $N_g$ as a function
of the free parameter in specific models listed in Table 1; the two 
curves correspond to the calculations for the two different epochs, 
the upper one showing case (i) whereas the lower one case (ii).
We also indicate as the thick lines the normalization $\sigma_8$ given 
by the 4-year COBE data as a function of $h$ in panel (a), and for
two choices of the Hubble constant, $h=0.5$ and 0.8, in 
panels (b)-(e). Since we take our abundance calculation
to give a lower limit on the halo abundance, the allowed range
lies in the lower-right region of the line indicating the COBE 
normalization. 

Let us now discuss each specific case given in Fig. 2. 
 Panel (a) shows the constraint on $h$ for the standard CDM models with
$\Omega_0=1$ and $n=1$. The allowed range is $h\gsim 0.35$ and
is not very sensitive to the changes of $\Delta t$. 
This limit is slightly higher than the upper
limit $h<0.3$ (so that $\Gamma\equiv \Omega_0h< 0.3 $) to give 
the required large scale clustering power at $z\sim 0$ obtained from the 
correlation function on scales near $10h^{-1}$ Mpc 
(e.g. Efstathiou, Sutherland \& Maddox 1990).
For the most standard case of $h=0.5$, the abundance 
limit gives $\sigma_8\gsim 0.6$ whereas the 
COBE normalization leads to $\sigma_8\sim 1.2$. Although
there is no conflict between the abundance limit and 
the COBE normalization, the gap between the two values of 
$\sigma_8$ implies that an order of magnitude more halos must 
have existed at $z>3$ without forming stars. 

  Panel (b) gives the constraint on the power index $n$ for CDM models 
with $\Omega_0=1$ and $h=0.5$. We obtain a limit $n>0.85$ for
$\Delta t=1$ Gyr [case (i)]. This limit is relaxed to
$n\gsim 0.75$ if we take $\Delta t=0.01$ Gyr [case (ii)].
To allow a value $n\sim 0.7$, we must take 
$h\sim 0.6$ and $\Delta t \ll 1$Gyr. On the other hand,
$n\sim 0.7$ and $h\sim 0.5$ is required to match the observations
on large scales at low redshift (e.g. Ostriker \& Cen 1996); 
such a model is, therefore, marginally ruled out 
from our abundance argument.
   
  Panel (c) shows the results for open CDM
models with $n=1$. The predicted abundance depends only  
weakly on $h$ and calculations are shown only for $h=0.5$.
The values of $\sigma_8$ for a given $N$
is almost flat against the change of $\Omega_0$.
The normalization given by the COBE data, however, depends strongly on
$\Omega_0$. We obtain a limit $\Omega_0\gsim 0.5$ for $h=0.5$ and
$\Omega_0\gsim 0.3$ for $h=0.8$. These limits are summarized
in terms of the shape parameter (of CDM-like power spectrum) 
as $\Gamma\gsim 0.25$. The fact that the abundance of $N_g$ is smaller and
the (linear) density perturbations grow slower in an open universe
makes the limit on $\Gamma$ lower than that obtained for the 
Einstein-de Sitter universe. The limit obtained here is marginally 
consistent with what is required to explain the large scale 
clustering power.

  Given in panel (d) is the constraint on $\Omega_0$ for flat CDM models
with $n=1$ ($\Omega_0+\lambda_0=1$). We present the abundance results 
for $h=0.7$, but the results depend only weakly on $h$.
The COBE data leads to a limit $\Omega_0\gsim 0.4$ for
$h=0.5$. For $h=0.8$, the limit is $\Omega_0\gsim 0.2$. 
In terms of $\Gamma$ the limit is $\Gamma\gsim 0.16$-0.2.
This range of $\Gamma$ (or $\Omega_0$) well overlaps with the
range derived from the clustering of galaxies on large scales. 
In particular, a flat CDM model with $\Omega_0\sim 0.3$ and 
$h\sim 0.7$, as favoured by Ostriker \& Steinhardt (1996), is 
perfectly consistent with the observed abundance of giant galaxies
at high redshifts.  

  The last panel of Fig.2 shows the constraint on $\Omega_\nu$
for MDM models with $h=0.5$ and $\Omega_0=1$. 
The MDM model with $\Omega_\nu=0.3$
has been found to be successful in predicting the clustering
properties of galaxies on large scales (e.g. Jing et al. 1993;
Klypin et al. 1993), but it was later found
that this model do not have large enough power on small scales to 
explain the total baryon mass observed in damped Ly$\alpha$
systems (Mo \& Miralda-Escude 1994; 
Kauffmann \& Charlot 1994; Ma \& Bertchinger 1994; Klypin et al.
1995). In panel (e) we see that the combined  COBE/abundance limit
gives much stronger constraint on $\Omega_\nu$: if
$h=0.5$ any MDM models with $\Omega_\nu\gsim 0.2$ are inconsistent
with this limit. Lower values for $\Omega_\nu$ 
are still allowed, but the advantage of the MDM models in
explaining large scale clustering power would be lost for 
such a small $\Omega_\nu$. 

\section {PREDICTION FOR THE CORRELATION FUNCTION}
  The bias parameter of dark halos, $b$, is defined by the ratio of
the two-point correlation function of halos $\xi$ to that of mass
$\xi_m$ as $\xi(r)=b^2\xi_m(r)$. Mo \& White (1996) 
argued that this bias parameter is a function of
velocity dispersion $\sigma$ at redshift $z$ and is accurately
described (to moderately nonlinear regime) by
$$
b(\sigma, z)=1+{1\over \delta_c(z)}
\left\lbrack {\delta_c^2(z)\over \Delta^2(r_0)}-1
\right\rbrack,
\eqno(2)
$$
where $r_0$ is specified by $\sigma$ as discussed in Section 2. This
$b$ parameter refers to the bias parameter of galaxies 
if galaxies were formed at the center of the halos and  
have not lost their identities during the subsequent evolution.

In Figure 3 we plot $\sigma_{8,g}\equiv \sigma_8b$ as a function 
of $\sigma_8$ for the `Lyman break galaxies' in flat CDM models, where
$b$ is the average of $b(\sigma,z)$ over $\sigma$ with a weight 
of $dN/d\sigma$.
According to the above interpretation, $\sigma_{8,g}$ is the rms 
fluctuation of counts of these `galaxies' in spheres of
radius $8h^{-1}$Mpc at present time. This value should be compared to that
for present-day normal galaxies, $\sigma_{8,g}\approx 1$ 
(e.g. Davis \& Peebles 1983). Fig.3 shows
that $\sigma_{8,g}$ is substantially larger than unity in all cases,
implying that `Lyman break galaxies' in these models are
significantly more strongly clustered than normal galaxies.
For $\Omega_0=0.3$ and $\sigma_8\sim 1$, $\sigma_{8,g}$ is about
1.8. Thus the amplitude of the correlation function of these galaxies
at present time should be about 3 times as large as that of
normal galaxies, or comparable to that of giant early-type galaxies
(e.g. Davis \& Geller 1976; Jing, Mo \& B\"orner 1991).
This prediction corroborates the arguments that the observed 
Lyman break galaxies are the progenitors of present-day
large E/S0 galaxies and is in agreement with the preliminary
observational result of Giavalisco et al. (1994). 

 \section {CONCLUSIONS}

 We have demonstrated that the abundance
of giant galaxies at high redshift gives significant constraints
on models of cosmic structure formation 
and discriminate among
models that satisfy other currently available tests.
Using COBE normalization of the density perturbation spectrum, 
we have shown that the abundance of Lyman break 
galaxies, as observed by S96, already rules out a number of 
current models of structure formation. In particular, 
the CDM models that are devised to give large
scale clustering power by lowering the Hubble constant or by tilting the
initial density power spectrum are disfavoured, 
leaving the case with a low density universe as  
marginally allowed. We are left with the moderately $\Lambda$-dominated
CDM models as the ones that satisfy all the constraints. 
We have also shown that the MDM models that can explain 
large scale clustering power are strongly disfavoured by the 
same argument. 

In this {\it Letter}  we have concentrated only on the tests based 
cosmic structure formation. There are, of course,  constraints 
from many other observations, 
such as those on the Hubble constant and on the age of the universe, 
and those from gravitational lensing observations,
that we have not discussed here.  
It is interesting to note that our favoured 
``moderately $\Lambda$-dominated CDM model'' 
is consistent with all available constraints. 

We have also argued that if cosmic large-scale structure forms 
in hierarchical clustering, high-redshift giant galaxies should be 
more strongly clustered at present time than normal galaxies, 
corroborating the interpretation that Lyman break galaxies are 
progenitors of early-type galaxies. 

One caveat is that our argument hinges on the assumption that the
velocity dispersion observed by S96 is gravitational, that is, the
galaxies they observed are giant galaxies. If the high velocities
are dominantly non-gravitational, the conclusions we derived
should all be modified, so is the interpretation given in S96.
We are very much looking forward to the confirmation of this point in
future high resolution spectroscopy.  When this doubt is cleared
up, even the current, rather premature observational data 
of high redshift galaxies can serve as an important diagnostics 
for models of cosmic structure formation.   

\acknowledgments
We are very grateful to Wayne Hu for his advice in COBE
normalizations, and to Chung-Pei Ma for providing us with her
MDM spectra and her advice in using them. We also thank
John Bahcall, Harry Ferguson, Jordi Miralda-Escud\'e, Martin Rees and
Alex Szalay for stimulating discussions. 
HJM is supported by the Ambrose Monell Foundation and MF acknowledge
the support from Fuji Xerox Corporation.

\begin{deluxetable}{lccccc}
%\tablewidth{33pc}
\tablewidth{0pc}
\tablecaption{Model Parameters}
\tablehead{
\colhead{model}       & \colhead{$h$} &
\colhead{$n$}           & \colhead{$\Omega_0$} &
\colhead{$\lambda_0$} & \colhead{$\Omega_{\nu}$}}

\startdata
standard CDM&varying     &1.0      &1.0            &0.0     &0.0  \nl
tilted CDM  &0.5         &varying  &1.0            &0.0     &0.0  \nl
open CDM    &0.5-0.8     &1.0      &varying        &0.0     &0.0  \nl
flat CDM    &0.5-0.8     &1.0      &$1-\lambda_0$  &varying &0.0  \nl
MDM         &0.5         &1.0      &1.0            &0.0     &varying \nl
\enddata
\end{deluxetable}

\begin{figure}
%\epsscale{1.0}
\plotone{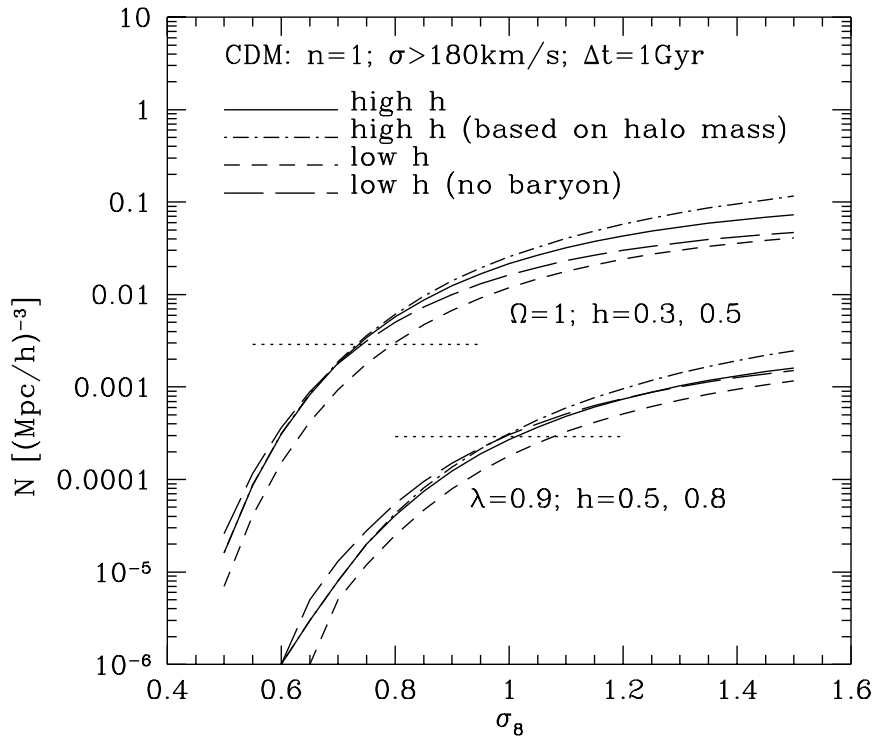}
\caption{Comoving number densities of halos at $\Delta t=1$Gyr
before the epoch of $z=3$ predicted by the standard
CDM model ($\Omega_0=1$) and by a CDM model in a flat universe
($\Omega_0=0.1$, $\lambda_0=0.9$). For each model results are
shown for two values of $h$ (other parameters are
fixed to be at their fiducial values), for one case where $\Omega_b$
is set to be zero (for the lower $h$ case only) and for a calculation
using halo mass (see text).  The horizontal dotted line shows the
observed abundance of Lyman break galaxies (S96) estimated
for the relevant cosmology.
}\end{figure}

\begin{figure}
%\epsscale{1.0}
\plotone{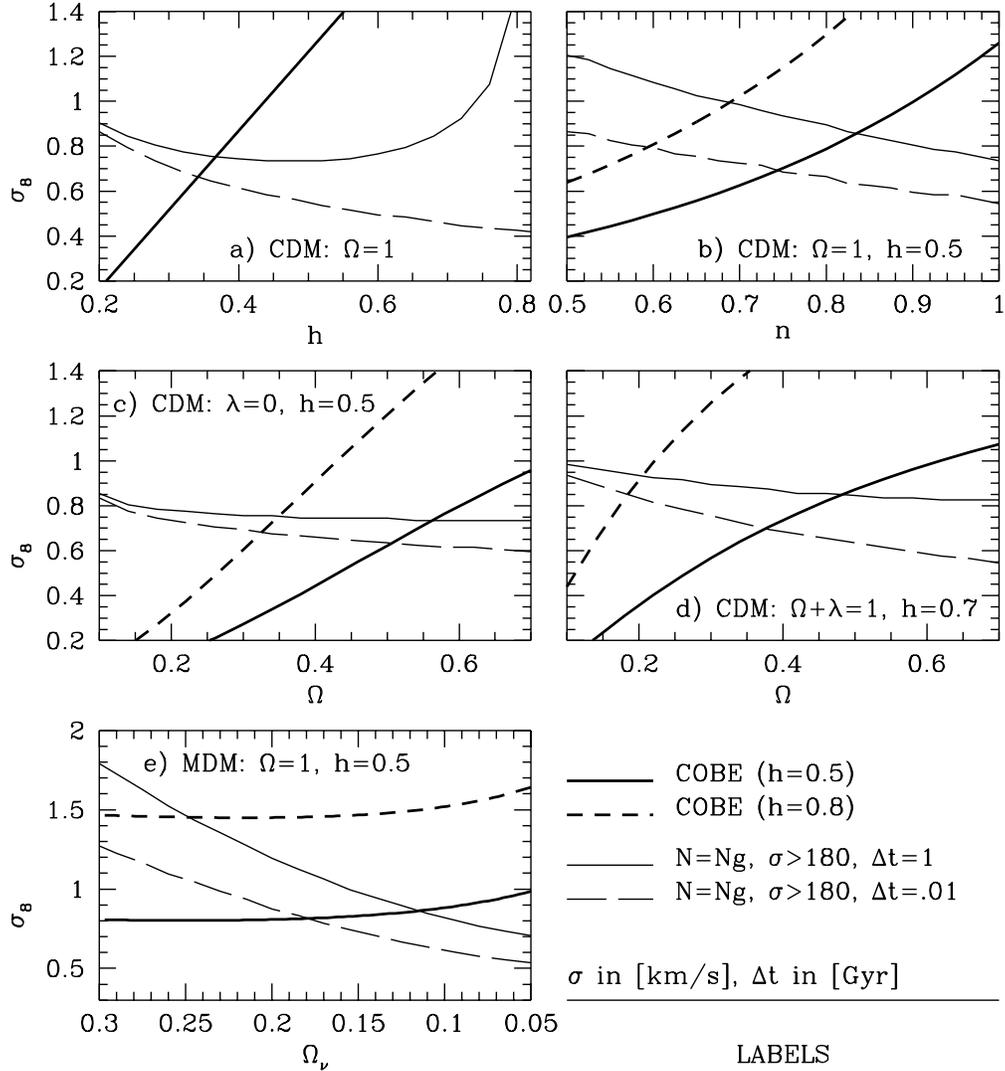}
\caption{
The values of $\sigma_8$ required to have the
predicted halo abundance to be equal to the observed abundance
of Lyman break galaxies ($N_g$).  The value of $\sigma_8$ given
by the 4-year COBE data are also shown by thick curves (when
two such curves are shown, they refer to two different $h$).
Since we require $N\ge N_g$ (see text), the allowed
region is below the curve indicating the COBE normalization
(thick curves). Panels (a)-(d) are for CDM models, and (e)
is for MDM models.
}\end{figure}

\begin{figure}
%\epsscale{1.0}
\plotone{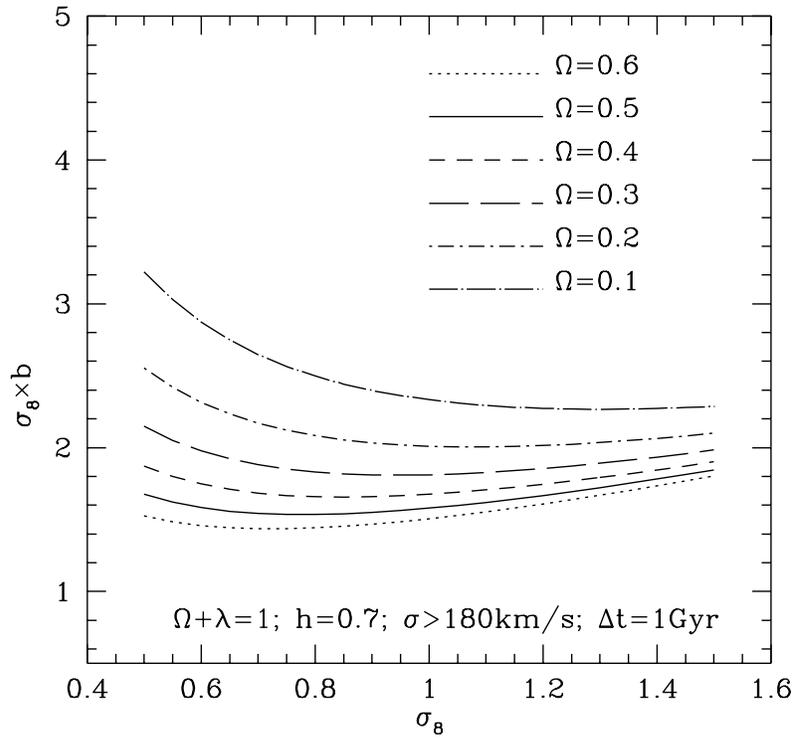}
\caption{The bias parameter $b$ (times $\sigma_8$) for the halos
of `Lyman break galaxies' in flat models with various $\Omega_0$ 
and $\sigma_8$. A value of $\sigma_8\times b>1$ means that these
`galaxies' are more strongly correlated than present day
normal galaxies.}\end{figure}
\end{document}